\documentstyle[11pt,aaspp4,flushrt]{article}  

\def\ref{\par\noindent\hangindent=1in\hangafter=1}
\def\etal{et al.}			
\def\H0{$H_0$~= 75 \kms\ Mpc$^{-1}$}

\def\kms{km s$^{-1}$}

\def\ref{\par\noindent\hangindent 30pt}

\def\farcs{\hbox{$.\!\!^{\prime\prime}$}}

\def\v16{$\Delta V_{1-6}$}

\begin{document}

\clearpage

\thispagestyle{empty}	


\begin{center}

\begin{Large}

{\bf WHAT FRACTION OF THE YOUNG CLUSTERS IN 
THE ANTENNAE GALAXIES ARE ``MISSING'' $^1$?} \\

\end{Large}

\end{center}

\begin{large}			

\bigskip
\bigskip

\centerline{BRADLEY C. WHITMORE  \& QING ZHANG$^2$}

\centerline{Space Telescope Science Institute, 3700 San Martin Drive,
Baltimore, MD 21218}
\centerline{Electronic mail: whitmore@stsci.edu}

\bigskip
\bigskip

Received:

\bigskip
\bigskip

\noindent $^{1}$ Based on observations with the NASA/ESA {\it Hubble Space 
Telescope}, obtained at the Space Telescope Science Institute, which is
operated by the Association of Universities for Research in Astronomy,
Inc., under NASA contract NAS5-26555.

\noindent $^{2}$ Also Johns Hopkins University, 3700 San Martin Drive,
Dept. of Physics \& Astronomy , Baltimore, MD, 21218.

\clearpage

\centerline{\bf ABSTRACT}

A reexamination of the correspondence between 6 cm radio continuum
sources and young star clusters in the
Antennae galaxies  indicates that $\sim$85 \%
of the strong thermal sources have optical counterparts, once the
optical image is shifted 1$\farcs$2 to the southwest. A sample of 37
radio-optical matches are studied in detail showing correlations
between radio properties (i.e., total flux and spectral index) and a
variety of optical characteristics (i.e., intrinsic cluster
brightness, H$\alpha$ flux and equivalent width, extinction, and
cluster ages).  There is a strong correlation between the radio flux
and the intrinsic optical brightness.  In particular, the brightest
radio source is also the intrinsically brightest optical cluster
(WS80). It is also the most extincted cluster in the sample, the
strongest CO source and the strongest 15
micron source .  Furthermore, the brightest ten radio sources are all
amongst the youngest clusters with ages in the range 0 - 4 Myr and
extinctions from A$_V$ = 0.5 to 7.6 mag (with a median value of 2.6
mag). A weak correlation between age and A$_V$ suggests that $\sim$6 Myr
are typically required to remove enough dust to reach A$_V$ = 1 mag.  The
radio-bright phase lasts only about 10 Myr in these clusters,
consistent with the interpretation that most of the radio emission
originates from hot gas.  Many of the non-detections are
probably individual supernova remnants since they have relatively
steep radio indices typical of non-thermal sources.  Only
a few of the very red clusters originally discovered by Whitmore \&
Schweizer are radio sources, contrary to earlier suggestions.
Finally, a new hybrid method of determining cluster ages has been
developed using both UBVI colors and H$\alpha$ equivalent widths to
break the age-reddening degeneracy. We find that the Bruzual \&
Charlot models, which use the Padova spectral evolution tracks, fit the
data reasonably well while the Leitherer \& Heckman models, using the
Geneva tracks, have a large red loop for ages 8 - 13 Myr that does not
agree with the data.

\bigskip


\clearpage

\centerline{\bf 1. INTRODUCTION}

Star formation requires an ample reservoir of cool molecular gas as
raw material. Along with the gas comes dust, resulting in the ironic
situation that many of the brightest stars cannot be seen in the
optical due to obscuration by dust. Fortunately, radio and infrared
observations are able to penetrate much of the dust, providing a
window into the star formation process. This is important for two
basic reasons: 1) it allows us to observe the stars using many
different wavelengths, thus providing better diagnostics for
unraveling the processes involved in star formation, 2) it allows us
to determine whether conclusions based on an incomplete census of
optical sources are jeopardized by incompleteness.

Merging galaxies are the sites of the most active star formation in
the universe.  The Antennae Galaxies (NGC 4038/39) represent perhaps
our best chance to study the process in detail, since they are  both the
youngest and nearest galaxy in Toomre's (1977) list of 11 prototypical
mergers. Whitmore \etal\/ (1999) have identified a rich population of
young compact star clusters in the Antennae, many of which have all
the attributes expected of young globular clusters. The ``overlap''
region between the two galaxies is the site of the most active star
and cluster formation in the system.  Unfortunately, the extensive
dust in this region makes it difficult to study the clusters in
detail.  Several authors have highlighted the close connection between
the clusters, infrared sources (Vigroux \etal\/ 1996), sources of CO
emission (Wilson \etal\/ 2000), and a variety of sources observed in
other wavelengths (Zhang, Fall, \& Whitmore 2001). However, other
authors have stressed the difficulties of penetrating the dust, and
have concluded that most of the sites of star formation may actually
be embedded in optically thick layers of dust with A$_V \sim$70 (Kunze
\etal\/ 1996, Mirabel \etal\/ 1998). For example, Neff and Ulvestad
(2000) state that ``The strongest radio continuum emission occurs
between the galaxies, at an optically unremarkable location near but
not coincident with an extremely red cluster detected by Whitmore \&
Schweizer.''

In this paper we reexamine the question of whether most of the strong
radio  sources are actually missing from the optical survey of
young clusters, due to extinction from dust. 
We  adopt a Hubble Constant of $H_0 = 75$ km s$^{-1}$
Mpc$^{-1}$, which places NGC 4038/4039 at a distance of 19.2 Mpc,
corresponding to a distance modulus of 31.41 mag.  
At this distance, the projected scale
is 1$''$ = 93 pc, and 1 pixel on the Planetary Camera covers 4.23 pc
while one pixel on the Wide Field Camera covers 9.26 pc.

\bigskip

\centerline{\bf 2. A COMPARISON BETWEEN RADIO AND OPTICAL POSITIONS}

\centerline{\it 2.1 Statistical Analysis}

Accurate coordinates are the primary requisite for the identification
of optical counterparts to radio and IR sources. This requires both
high spatial resolution, in order to pinpoint a given target, and an
accurate absolute reference frame shared by the two observations.  The
spatial resolution for the available IR observations in the Antennae
is roughly 2$''$ (Mirabel \etal\/ 1998). While this is good enough to
establish the fact that many of the IR sources have optical
counterparts (Vigroux \etal\/ 1996, Mirabel \etal\/ 1998), it is difficult to make precise comparisons since
there are often several optical sources within a search radius. The radio
observations from Neff \& Ulvestad (2000) provide much
better positional accuracies, with uncertainties $\sim0\farcs$4,
based on a comparison of their 6 and 4 cm positions for the same
cluster (note that the 6 cm positions are used for comparisons in the
present paper).  HST observations have {\it relative} accuracies
$\sim$0.1$''$ (Voit 1997).  Unfortunately, the absolute
uncertainties of the HST positions are much larger, typically
$\sim$1$''$ (Biretta \etal\/ 2001), due to uncertainties in the absolute positions of the
guide stars.  Hence, an offset of 1 - 2$''$ should be considered when
making positional matches with HST data.

Figure 1 {\it (bottom)} shows the positions of the nine bright (i.e., radio flux S
$\geq$ 70 $\mu$Jy), thermal (i.e., radio index $\alpha \geq -0.4$, where
$\alpha$ is defined by S$_{\nu} \propto \nu^{+\alpha}$), 6 cm
radio sources that fall in the overlap region of the Antennae (Neff \&
Ulvestad 2000, Table 5).  We begin our comparison using this subset of
the 109 radio sources because they are the most likely to have optical
counterparts, based on the fact that similar sources in the Milky Way
are often associated with compact HII regions.

Neff \& Ulvestad (2000) found that many of the radio sources are near,
but not coincident with optical sources. They suggest that the true
counterparts are embedded in dust, and are not observed in the optical
passbands. However, we note from Figure 1 {\it (bottom)} that the nearest optical
candidates are generally seen toward the northeast, hence a common
offset may bring the radio and optical positions into agreement. The
result of adding a 1$\farcs$2 offset toward the southwest is shown in
Figure {\it (top)}.

We now
find that eight of the nine radio sources have optical counterparts
(defined as point-like sources from Whitmore \etal\/ 1999 with M$_I$
$<$ --9.0) which are within 0$\farcs$5 of the radio positions. Without
the offset, only three of the radio sources would have matches given
the same criteria, which is roughly what would be expected for a
random distribution (see Figure 2 and discussion below). 

If we expand the field of view to include the full HST image we find a
similar story, with matches for 11 of the 13 bright thermal radio
sources using the 1$\farcs$2 offset. Using no offset yields only 3 of 13
matches. The chances of matching 11 out of 13 objects is 5.1 $\times$ 10$^{-12}$,
assuming a random distribution, a probability of 0.0641 for a single
trial (based on a search radius of 0$\farcs$5 and the total field of view
of the WFPC2), and using a binomial probability distribution (i.e.,
[n!$\times$(n-m)$\times$p$^{m(1-p)}$]/[(n-m)!$\times$m!]; where n is the
number of trials, m is the number of matches, and p is the probability
of a match in a single trial). However, this approach underestimates
the probability since the sources are not distributed randomly over
the full field of view. A more realistic estimate can be made by
performing Monte-Carlo calculations by offsetting the optical sources
by small spatial offsets (i.e., using a grid with delta-X and delta-Y
in the range $\pm$ 5$''$, with 1$''$ increments), and then searching
for matches. This preserves the large-scale distribution of the
sources. Figure 2 shows that the mean number of predicted random
matches is 2.18, which translates to a probability of 0.168 for a
single trial. The curve in Figure 2 shows the predicted probability
using the binomial formula and P=0.168. It is in good agreement with
the results from the Monte-Carlo calculation, giving us confidence in
an extrapolation to larger numbers of matches than is practical to
obtain using the Monte-Carlo calculations. The predicted probability
for 11 matches for 13 objects is 4.6 $\times$ 10$^{-7}$, according to the binomial
formula. This clearly demonstrates that the 1$\farcs$2 offset is
justified, and there are optical counterparts for nearly all of the
bright thermal radio sources.

The only bright thermal radio sources that do not have matches are 1-3
and 4-4 (radio designations are from Table 5 of Neff \& Ulvestad 2000). 1-3 is in a
dust lane 5$''$ NW of the nucleus of NGC 4039.  The radio source is relatively
bright (S=303 $\mu$Jy) and is atypical in the sense that it has
the highest value of the radio index $\alpha$ (+0.38) of the 13 galaxies. 
Radio source 4-4 is in a dust lane halfway
between regions C and D (see Figure 5 of Whitmore \etal\/ 1999), and
can be seen as the circle closest to the North arrow in Figure 1 {\it 
(bottom)}. It
is also quite bright in the radio (S=680 $\mu$Jy) and has a typical value of $\alpha$
(--0.16). Taken at face value, this suggests that 2/13 = 15 \% of the
bright thermal radio sources are embedded in sufficient dust to
obscure their optical emission in the HST observations.

It will be important to observe the two undetected radio sources using high 
resolution IR observations, to determine whether they are obscurred
star clusters, and if so, whether they have properties
similar to the detected radio sources. For example, if these clusters
turn out to be amongst the intrinsically brightest clusters, it is possible
that they may influence the  luminosity function in important ways,
even though there are relatively few missing clusters.

Similar enhancements in the number of matches are found for other
subsets of the radio sample once the 1$\farcs$2 offset is made. These
subsets are listed in Table 1.  In particular, 77\% of the 13 very
bright radio sources (S $>$ 300 $\mu$Jy) have optical counterparts,
and 62\% of the 45 bright sources (S$\geq$ 70 $\mu$Jy) have optical
counterparts. However, when only the faint radio sources are used (S
$<$ 70 $\mu$Jy), there is essentially no enhancement in the number of
counterparts. This may indicate that the radio sources are associated
with sources other than star clusters. For example, it is possible
that they arise from individual supernova remnants, which may not
appear bright enough to be included by the M$_I$ $<$ --9 selection
criteria.

We note that only $\sim$10 \% of the apparent $U$ light from the
Antennae stems from clusters (i.e., the fractions are 9/8/5/7 \% for
U/B/V/I respectively; the cluster-rich PC has the highest percentage
of light in clusters with 16/22/15/21~\%; the WF4 has the lowest with
4/3/2/3~\%; and the overlap region is intermediate with 9/8/5/7~\%).
Similar fractions were found in NGC 3256 by Zepf et al. (1999).  These
fractions should be considered lower limits, since only clusters
brighter than M$_{I}$ = --9 were included in the calculation. In addition,
we assume that the extinction is the same in the clusters and in the field.
Nevertheless, it appears that a majority of the $U$ light comes from
field stars rather than clusters.  This might explain why most of the
faint radio sources, apparently associated with individual supernova
remnants, appear to be associated with the field rather than the
clusters.  We should also keep in mind that some of the weakest radio
sources are likely to be noise rather than real sources, hence
artificially reducing the number of matches for the faint sources.

Table 2 includes the 37 radio sources with good positional matches
with the M$_I$ $<$ --9 clusters (within 0$\farcs$5), once the 1$\farcs$2
offset is made. A Monte-Carlo calculation indicates that 13.8 matches
would be expected if there were no correlation between radio and
optical sources.  This indicates that p=0.1265 for a single trial. 
The chances of having 37 matches from the 109 radio sources is 6.0 $\times$
10$^{-9}$, according to the binomial formula.
Hence, based on statistics alone,  we believe there is compelling
evidence for an offset
of 1$\farcs$2 between the radio and optical image.

\bigskip

\centerline {\it 2.2 Astrometric Analysis}

Offsets of 1$''$ - 2$''$
are often seen in HST images, due to uncertainties in the guide star
positions. Hence, in cases in which accurate astrometric positions
are required, observers are urged to 
check for offsets in the WFPC2 positions by measuring the
position based on measurements from the Digital Sky Survey (DSS),
following the procedure described at:
http://www-gsss.stsci.edu/support/phase2.html .

Unfortunately, the ground-based DSS image of the Antennae has only a
few point-like objects in common with our HST images, since most
of the DSS image is saturated. However, it is possible
to measure a region to the NE of region D (see Figure 5a in Whitmore
\etal\/ 1999) on both the DSS image and the WFPC2
image. Using this cluster we find an offset of --0.036~s (or
-0$\farcs$51) in RA and --1$\farcs$05 in DEC.
This is similar to
our empirically determined values of --0.035~s (or --0$\farcs$49) in RA
and --1$\farcs$07 in DEC discussed in \S 2.1 .  Expected accuracies for the
DSS coordinates are $\sim$0$\farcs$2 -- 0$\farcs$3.

A final check is possible using a star on WF2 (i.e.,
star 2 in Figure 5b of Whitmore \etal\/ 1999), which turns out to be
an USNO astrometric reference (i.e., obtained from their archival web
site at http://www.nofs.navy.mil/data/FchPix/). 
We find that the offset it implies for the
HST images it is in very good agreement with
our previous estimates (i.e., --0.033 s [or -0.47$''$] in RA and
--1.16$''$ in DEC).

Hence, there is good evidence from both the statistics discussed in \S
2.1, and the independent astrometric checks discussed in this section, that an offset
with an amplitude of $\sim$1$\farcs$2 toward the SW is required to bring the WFPC2
image into alignment with the radio image.  The good
correlations between the radio and optical properties, 
discussed in \S 5, provide a final piece of confirming evidence
for this offset.

\bigskip

\centerline{\bf 3. IDENTIFICATION OF RADIO AND OPTICAL COUNTERPARTS}

Having established that an offset of the HST coordinates 1$\farcs$2
to the southwest is required to bring the radio and optical
images into alignment, we now attempt to identify
specific matches.
Figure 3 shows the locations of all 109 radio sources. The
large circles show matches with bright thermal sources (S $\geq$ 70 $\mu$Jy, $\alpha \geq$
--0.4) while the large squares show matches with bright non-thermal sources
(S $\geq$ 70 $\mu$Jy, $\alpha <$ --0.4).  The small symbols are for the 
faint radio sources (i.e., S $<$ 70 $\mu$Jy).

Each of the 37 matches was inspected in order to select the most
likely optical counterpart, since there are many cases where more than
one cluster is found within the search radius of 0$\farcs$5 from the
radio position.  During this process we noted a tendency for some of
the radio source to be closer to strong H$\alpha$ emission peaks than
to bright sources which are not associated with H$\alpha$.  Perhaps
the best example is Knot S (near the top of Figure 4) which is
optically the second brightest (apparent) cluster in the Antennae.
The radio source does not appear to be associated with the cluster
itself, but instead is found $\sim$1$''$ to the southeast, closer to
several regions of strong H$\alpha$.  In these cases we identified two
possible counterparts, one where the brightest cluster in the circle
was generally selected (in some cases a slightly fainter object was
selected if it was much closer to the exact radio position; see
Appendix A), and one where the strongest H$\alpha$ source was
chosen. In some cases a strong H$\alpha$ source just outside the
search circle was selected.  We will call the first sample the
brightness-selected sample and the second the H$\alpha$-selected
sample.  A third sample, consisting of the 37 matches discussed
above but using the total  H$\alpha$ flux within a radius of
of 0.5$''$ (i.e., the ``radio aperture''), has also been defined.
This will be called the radio-position sample. 
In \S 5 we attempt to identify which is the more physically
meaningful sample by examining the resultant scatter in various
correlations. The three samples, along with their radio counterparts and
other ancillary information, are listed in Table 3.

While the good correlations we find in \S 5 indicate that
most of these radio-optical matches are probably correct, it should
be kept in mind that based on statistics alone, some of the matches 
are likely to be misidentifications. Hence, caution is advised when making
specific one-to-one matches. We also note that the intrinsic
correlations will be better than indicated in \S 5, since the misidentifcations
will add noise.

\bigskip

\centerline{\bf 4. AGE DATING THE CLUSTERS}

In \S 5 we will compare various radio and optical properties of the 37
radio sources with optical counterparts.  In the present section we
first fine-tune our age estimates for the clusters, since several
properties we will use in these comparisons are derived from these
determinations.

Whitmore \etal\/ (1999) estimated ages for the clusters in the
Antennae by comparing the UBVI colors with Bruzual \& Charlot
(1996) spectral evolution models. They found evidence for five
populations of clusters, ranging in age from $\sim$1 Myr to $\sim$15
Gyr.  The middle panel of Figure 5 shows a $U-B$ vs $V-I$ diagram for the 37 radio sources
with optical counterparts, along with the Bruzual \& Charlot (2000)
instantaneous burst spectral evolution models with solar metallicity
(using their models with theoretical isochrones)
as the solid line, and Starburst99 (Leitherer \etal\/, 1999) models as
a dotted line. 
The models  are shown alone in the top panel, along with the Bruzual
\& Charlot (2000) models  with empirical isochrones for comparison (dot-dashed lines). 
The positions for ages of 1, 5, 10, 100, and 1000 Myr for the BC00-theoretical
models are shown in large numbers while the positions for 1, 6.5, 8 and 13
Myr for the Starburst99 models are shown in smaller numbers. 
The straight dashed lines show the reddening vectors due to
obscuration by dust  (Mathis 1990) for ages of 1, 5 and
10 Myr and the BC00 models with theoretical isochrones.

Some of the  large differences between the Starburst99 models
and the Bruzual \& Charlot models are due to the dominating influence
of cool red supergiants in the Geneva spectral evolution models used
by Starburst99, relative to the Padova models used in the
Bruzual \& Charlot models.  We note that of 37 objects, only cluster 984
lies in a position that would populate the prominent ``red loop'' (i.e.,
ages 8 - 13 Myr) seen in the Starburst99 models.  The bottom
panel of Figure 5 shows the distribution for the 100 
(apparent) brightest clusters in the Antennae, where no clusters would populate the red
loop. The Bruzual \& Charlot models with theoretical isochrones appear
to provide a better fit to the majority of the data, and hence will be
used in the remainder of the paper.
We note that a similar effect can be seen for the clusters in M83
(Harris \etal\/ 2002; Figure 6).

Object 841 is the nucleus of NGC 4039, and sits in a position
that would imply an age of $\sim$1 Gyr. This is likely to be
caused by a combination of light from an old population ($\sim$15 Gyr), intrinsic
to the pre-existing nucleus, and a
young population of stars formed during the merger. The other clusters in this
region of the diagram (3816, 2560, \& 3475) appear to be normal star
clusters. The most likely explanation for their position in the 
diagram is that they are young clusters
heavily embedded in dust, since all three are found in the heart of the 
overlap region, where extensive dust is clearly apparent. This interpretation
is strengthened by the fact that all three have relatively strong
H$\alpha$ emission, implying ages $<$ 10 Myr.   We note that this would require  
the true reddening
vector to be slightly steeper than the Mathis (1990) models used in the diagram.
Finally, cluster 1139 (=WS80) is so red that it suffers from a severe red leak
in the U filter, which artificially enhances the U magnitude. This is further
discussed below.

\bigskip

{\it 4.1 Breaking the Age-Reddening Degeneracy}

Our basic problem in determining ages is that for most clusters with
A$_V$ greater than about 1 mag, the reddening vector for a particular
cluster intersects the models in two, or even three points,
corresponding to two or three possible ages. For example, cluster 7453
has three possible ages; $\sim$1 , $\sim$7, and $\sim$40 Myr,
depending on how far up the reddening vector we need to
backtrack. Each of the three ages would also imply a different value
of A$_V$.

Fortunately, we can break this degeneracy by using the strength of
H$\alpha$ emission.  Figure 6 shows the predicted values of the
H$\alpha$ equivalent width (EW, in \AA) as a function of age, based on
solar metallicity, an instantaneous burst, and a Salpeter initial mass function
extending from 1.0 to 100 M$_{\odot}$ (Starburst99 models in Leitherer \etal\/, 1999).
For ages less than 4.8 Myr (where the curves in the $U-B$ vs
$V-I$ diagram take a sharp turn as the red giants begin to dominate,
resulting in the first level of degeneracy), the clusters should have
log H$\alpha$ $>$ 2.65, according to the models.  By 8.7 Myr, where
the next bend appears in the $U-B$ vs $V-I$ diagram resulting in the
second level of degeneracy, log H$\alpha$ has dropped to 1.2. This is
lower than the observed value in 36 of the 37 radio sources with optical counterparts, the
only exception being object 841, which is probably the nucleus of NGC
4039. This object has been removed from the subsequent
analysis. Hence, we can use the log H$\alpha$ = 2.65 criterion to
identify whether the cluster belongs in the 1 -- 5 or the 5 -- 9 Myr age
range. None of the radio-optical matches appear to have larger ages.
Once the correct age is known we use the difference between the
apparent and intrinsic values of $V - I$ to determine the extinction, A$_V$.

Values of H$\alpha$ have been measured using the F685N observations
from Whitmore \etal\/ (1999).  The H$\alpha$ flux (using a conversion
value of 4.2 $\times$ 10$^{-15}$ erg/s/cm$^2$ for 1 DN/s; from
http://www.stsci.edu/instruments/wfpc2/Wfpc2$\_$faq/wfpc2$\_$nrw$\_$phot$\_$faq.html)
was determined using a scaled version of the I band image to subtract
the continuum. Although the offset in wavelength between H$\alpha$ and
the I band is non-optimal, this technique works well enough for our
purposes since H$\alpha$ is a steep function of age (see Figure 6), so
small uncertainties can be tolerated.  The I-band image is then used
to estimate the value of the continuum and convert the H$\alpha$ flux
to equivalent width, using the values of photflam from Table 28.1 of
the Data Handbook (Voit 1997). A small correction is made for the difference
in extinction between H$\alpha$ and the I-band using Mathis (1990).
Figure 6 shows that the data follow the predicted curve relatively
well.

{\it 4.2 Correcting the Colors for the Presence of H$\alpha$}

Several practical issues need to be considered before final
values for the ages can be determined. The first is the fact that the
$V$ observations are contaminated by the presence of strong hydrogen
(4861 \AA) and oxygen emission lines (4959 and 5007 \AA), as
demonstrated by Stiavelli \etal\/ (1998) and also shown in Figure 16
of Whitmore \etal\/ (1999).  We have determined the correction for
this effect by using the $B-V$ vs $V-I$ diagram, since the age and
reddening vectors are nearly parallel for this combination of
colors. This simplifies the problem by allowing us to solve for just
two parameters rather than three (i.e.,
age, reddening, and contamination from emission lines). Figure 7 shows the raw $B-V$ vs $V-I$
diagram, with the ten clusters with the highest values of log
H$\alpha$ (used as a surrogate for  emission in the 4861, 4959, and 5007 \AA\/
lines)
shown as open circles. As expected, these points tend to
fall to the left of the models.  Figure 8 shows the residuals from the
1 Myr BC00 models vs.  log H$\alpha$. We choose the 1 Myr models for
reference since the clusters with large values of log H$\alpha$ are
all very young (see Figure 6). Also note that the reddening lines for
all ages less than 5 Myr are nearly identical in the $B-V$ vs $V-I$
diagram.

We correct for the log H$\alpha$ dependence of $V-I$ using the
equation:

\noindent $(V-I)_{cor}$ = $(V-I)_{apparent}$ + 2.5 $\times$ log [(1.4 $\times$ 10$^{-4}$ $\times$ H$\alpha$) + 1]  $\phantom{aaaaaaaaaaaaaaaaaa}$ (1)

and correct for the dependence of $B-V$ using the equation

\noindent $(B-V)_{cor}$ = $(B-V)_{apparent}$ --2.5 $\times$ log [(1.1 $\times$ 10$^{-4}$ $\times$ H$\alpha$) + 1],  $\phantom{aaaaaaaaaaaaaaaaaa}$ (2)

where H$\alpha$ is measured as an equivalent width in \AA.

These formulae were obtained by zeroing out the mean value of the
residual for the ten clusters with the highest values of
H$\alpha$. Figure 8 shows both the correlation and the resulting
correction (dashed line) for $B-V$. Figure 9d shows the resulting $B-V$ vs
$V-I$(cor) diagram after the correction has been made.

Another difficulty is introduced by the very red clusters, where 
large extrapolations are required. Any small uncertainty in the
reddening law therefore results in large uncertainties in the age
estimate. In addition, for extremely red objects, the red leak in the
F336W filter becomes a problem (see Zhang, Fall, \& Whitmore 2001 for
a discussion). This is why cluster 1139, with $V-I$ = 2.92, has a value
of $U-B$ which is completely off-scale when compared to its value of
$V-I$ (see Figure 5). The U measurement appears to be $\sim$2 magnitudes too
bright, indicating that $\sim$85 \% of the light is coming from
the red part of the spectrum rather than the UV light. 
Cluster 984 may also suffer from a redleak problem, which 
is partly compensated by the fact that this cluster has the highest value of 
H$\alpha$, making the cluster appear bluer in $V-I$ than
it should. 

We deal with both of these problems by
making age estimates based on the color-color diagram only for
clusters with $V-I$ $<$ 1.2. For redder clusters we
make a rough age measurement based on the strength of H$\alpha$
emission alone.

\bigskip

\centerline{\it 4.3 Combining Age Estimates}

Combining the age estimates based on the three different color-color
diagrams is also problematic, since the uncertainties are often
extremely non-linear (e.g., the 1 and 3 Myr reddening lines differ by
only 0.04 mag in $V-I$ at a constant $U-B$ while the 7.2 and 8.7 Myr
reddening lines differ by 0.51 mag). Table 3 includes a subjective
``best guess'' age estimate, based primarily on the $U-B$ vs $V-I$
diagram since it provides the most discriminating power, and using
H$\alpha$ to break the age-reddening degeneracy. Lesser weight is
given to the other two color-color diagrams.  For clusters that fall
outside the possible age ranges predicted by the models (e.g., WS80 =
1139) we make rough age estimates based on H$\alpha$. A subjective
quality rating, reflecting the consistency and availability of the
various estimates, is included in Table 3.
Figure 9 shows all four of the diagrams used to make the best guess
estimate.  The dashed lines show the reddening vectors for 1, 4.8, 
and 8.7 Myr.

A final caveat is suggested by the fact that the H$\alpha$
measurements are made using the same apertures as the original UBVI
measurements (i.e., a radius of 2 pixels for the PC and 1.5 pixels for
the WF). However, for older clusters which have had time to blow
superbubbles, most of the H$\alpha$ emission associated with the
cluster may be at larger radii (e.g., Knot S in Figure 4 has very
little H$\alpha$ within a 2 pixel radius). Hence, the H$\alpha$
measurements should be regarded as lower limits. This does not appear
to be a serious problem, however, since nearly all of the radio
sources turn out to be associated with very young clusters where most
of the H$\alpha$ emission is still roughly coincident with the optical
source.

\bigskip

\centerline{\bf 5. CORRELATIONS BETWEEN RADIO AND OPTICAL PROPERTIES}

Having established that most of the strong 6 cm radio sources have optical
counterparts, we are now able to study the sources in more detail.
Figure 10 shows various correlations between radio and
optical parameters.  The significance of the correlation is included
in the upper right of each diagram. Obvious outlyers (labeled in Figure 10)
have been removed
before making the fits for  Figures 10d and 10g. The eight sources with  radio index $\alpha$
$<$ -0.8 (i.e., non-thermal) are shown as open symbols.
These sources generally follow the trends shown by objects with
flatter radio indices, although two  clusters (3816 and 10808)
stand out as outlyers in the log H$\alpha$ -- Av correlation (Figure 10g).
We draw four conclusions from these diagrams.

First: There is a weak tendency for young clusters to be stronger radio sources (Figures 10a, 10b).
This is not surprising since the UV radiation heating the
gas is at its maximum for very young stars, as is the H$\alpha$ EW  
which is controlled by the UV flux (Figure 6). This results
in the good correlation between log H$\alpha$ and 6 cm radio continuum emission (Figure 10b),
which has been well established by a number of past studies
(e.g., Young \etal\/ 1996).  

We note the fan shaped correlation between log H$\alpha$ and log S in
Figure 10b (and in other diagrams such as 10f and 10g). This may
indicate that the strongest radio sources are dominated by
the single cluster that has been matched with it, while weaker radio
sources may be associated with a different cluster, or
perhaps with a combination of several clusters, hence adding noise to
the correlation. This is discussed in another context at the end
of this section.

Second: Based on Figure 10d, the radio sources with optical counterparts appear to be primarily associated with 
objects with  $\alpha \geq$ --0.4, (i.e., thermal radio sources, Neff \& Ulvestad 2000).  
This is demonstrated by the constant upper envelope consistent with
a flat spectral index $\alpha$ $\approx$ -0.3.
This is the same conclusion as reached in
\S 2 based purely on statistical grounds.
There is little or no correlation between the
spectral index $\alpha$ and any other parameters.  Most of the
scatter occurs toward steeper radio indices for fainter
radio sources. These may represent a population of
sources associated with supernova remnants, or may just represent
misidentifications (i.e., noise). We also note that the source
with the largest value of $\alpha$ (cluster 5047 with $\alpha$ = 1.0) is also  the faintest
radio source, hence this may represent noise. 
This outlyer was not included when making the fit to the
data in Figure 10d.

Third: The younger clusters have the largest extinctions (Figure 10c,
10g).  Based on the weak correlation shown in Figure 10c, it takes
$\sim$ 6 Myr for a cluster to clear enough dust to reach A$_V$
$\sim$1.  This is consistent with the trend found by Zhang, Fall, \&
Whitmore (2001), with a mean value of A$_V$ $\sim$1.5 for the youngest
B1 clusters ($\leq$ 10 Myr) and A$_V$ $\sim$0.3 for the older B2
clusters ($\geq$ 100 Myr).  We note that while we expect a good
correlation between log H$\alpha$ and log Age, Figure 10f cannot be
used to independently prove this since it is largely built into the
age-dating method (see \S 4).

Fourth: The strongest radio sources are also the intrinsically brightest clusters
(Figure 10e). This is to be expected since the brightest optical
clusters are likely to be the strongest UV sources, due to a
combination of their relative youth (Figure 10f, 10h) and their higher
luminosity. The weak trend between M$_V$ and  log Age (Figure 10h) is due to the dimming
of the clusters with time, as predicted by the BC00 models (e.g., see
Whitmore \etal\/ 1997, Figure 13).

We can attempt to determine whether the radio sources are physically
related to individual bright clusters, individual bright HII regions,
or the mean H$\alpha$ within a 0$\farcs$5 radius of the radio source
by examining the degree of scatter in the various diagrams shown in
Figure 10.  Figure 11 shows the log S vs. log H$\alpha$ diagram for
all three samples defined in Table 3.  (i.e., the brightness-selected,
H$\alpha$-selected, and radio-position samples).  There is
essentially no difference between the brightness-selected sample and
the H$\alpha$-selected sample. This is also true of the other seven
correlations shown in Figure 10. However, the radio-position sample
does show a difference, with a flatter slope and considerably less
scatter (i.e., the RMS in log H$\alpha$ is 0.28, compared to 0.51 for
the brightness-selected sample and 0.53 for the H$\alpha$-selected
sample).  The steeper correlation for the bottom two panels in Figure
11 is probably due to misidentifications of objects with clusters with
weak H$\alpha$. From the top panel it would appear that essentially
all of the radio sources with optical counterparts are strong
H$\alpha$ sources.

While the measurement of log H$\alpha$ in units of equivalent width is
useful for our hybrid age determinations, the near constancy of log
H$\alpha$ shown in Figure 11c, and the strong correlation between log
S and M$_V$ (figure 10e), suggest that the underlying correlation may
be with the total H$\alpha$ flux within the ``radio aperture''.
Indeed, Figure 12 supports this interpretation, showing that the
correlation between log H$\alpha$ (flux) and log S for the
radio-position sample is stronger
(7.9~$\sigma$) than between any of the other parameters for
any of the three samples.

\bigskip

\centerline {\bf 6. WS80 - The Brightest Cluster in the Antennae}

Whitmore \etal\/ (1999) list Knot G (605 in Table 1 of Whitmore
\etal\/ 1999; see Figure 5 of that paper for location) as the
brightest cluster in the Antennae, based on its apparent value of $V$
(i.e., M$_V$ = -13.92, uncorrected for extinction).  We note that this
cluster is not a radio source, in agreement with its weak H$\alpha$
(see Figure 4 of Whitmore \etal\/ 1999). Mengel \etal\/ (2001)
estimate an age of 8 Myr for Knot G, older than our estimated ages for
all but one of the radio-optical matches in Table 3.

However, now that we can determine values of A$_V$ using the method
described in \S 4, we find that cluster 1139 ($\#$ 3 in Tables 2 and
3) is the intrinsically brightest cluster in the Antennae, with M$_V$
= --15.5, and a photometrically-determined mass of 4 $\times$ 10$^6$
M$_{\odot}$ (i.e., assuming the values for luminosity and age from
Table 2 and a BC00-theoretical model with solar metallicity; see Table 3
for the masses of other clusters).  This
cluster was originally identified by Whitmore \& Schweizer (1995) as
WS80, one of the very red objects that they suggested might be a young
cluster still embedded in its dust cocoon.

Since then, WS80 has also been identified as the strongest CO source
in NGC 4038/39 (Wilson \etal\/ 2000), the strongest ISO source
(Vigroux \etal\/ 1996, Mirabel \etal\/ 1998), and the strongest radio
source (Neff \& Ulvestad 2000, although they did not include the
1$\farcs$2 offset discussed in \S 2, and hence concluded that the true
radio source was near but not coincident with WS80). Wilson \etal\/
(2000) find that this is also the site of an apparent collision
between three giant molecular clouds.

WS80 is distinguished in our sample of 37 radio-optical matches as the
cluster with the highest value of A$_V$ (7.6 mag), the reddest color
($V-I$ = 2.92, or 3.08 after using Equation 1), the third highest
value of log H$\alpha$ (3.81), and one of the youngest ages ($\sim$2
Myr).

Figure 13 shows $U, B, V, I, V/I,$ and H$\alpha$ images of WS80 and
WS355, two very red clusters identified at the end of Table 1 in
Whitmore \& Schweizer (1995).  In both cases the clusters are nearly
invisible in the U and B images, are faint in the $V$ image, and are
bright in the I image.  This is in contrast to the blue clusters
around them which are roughly the same brightness in all four
colors. The contrast for the V/I image is adjusted to show the blue
objects as black and the red objects as white. WS80 is the only red
object in its field while there are two red objects in the WS355
field.  Note the elongation and the knots of nebulosity around
WS80. Most importantly, we find that WS80 is a strong H$\alpha$
source, with the knots aligned with the red knots in the V/I
diagram. In contrast, WS355 has essentially no H$\alpha$ associated
with it. This is presumably a somewhat older cluster which happens to
have several magnitudes of extinction in front of it, hence we would
not expect it to be a strong radio source.  Mengel \etal\/ (2001)
estimated an age of 8.5 Myr for WS355, consistent with its lack of
H$\alpha$ emission (i.e., see Figure 6).

The identification of WS80 with the strongest radio, CO, and IR
sources led to the expectation that many of the other very red sources
may be similar objects (e.g., Wilson \etal\/ 2000). However, we find
no other object in Table 3 have $(V-I)_{cor}$ $>$ 2.0. In addition,
when we isolate the fifteen optical clusters in the overlap region with $V-I$
$>$ 2.0, and M$_V$ $<$ --9, we find that the only match with a radio
source is WS80.  If we instead isolate the eleven intrinsically brightest 
optical clusters
in the overlap region with log H$\alpha$ $>$ 3, we find 8 of the 11
clusters are radio sources (and two more would be if we allowed a
1$''$ radius criterion for radio-optical matches).  Hence, it appears
that the primary criteria required to produce strong radio sources are the
intrinsic brightness of the cluster and the strength of its H$\alpha$
emission. Very red clusters are $\it not$ preferentially radio
sources, contrary to earlier suggestions.

\bigskip

\centerline{\bf 7. SUMMARY}

A reexamination of the spatial correspondence between 6 cm continuum radio
sources from Neff \& Ulvestad (2000) and young clusters in the Antennae
galaxies (Whitmore \etal\/ 1999) leads to the following conclusions.

1. The HST image used in Whitmore \etal\/ (1999) needs to be offset
1$\farcs$2 toward the southwest to bring it into alignment with the
radio image. Determinations based both on maximizing the number of
matches, and on a comparisons with independently determined
positions of objects on the image give similar results. 
Offsets of this magnitude are not uncommon in HST images, 
due to uncertainties in the coordinate system of the guide
stars. After the offset is made, 37 of the 109 radio sources have
optical counterparts. The probability of this occurring from an
uncorrelated sample is 6.0 $\times$ 10$^{-9}$.

2. Eight-five percent (11/13) of the strong thermal radio sources have
optical counterparts, indicating that only $\sim$15 \% of the sources
are embedded in so much dust that they are not detected in the
optical. Similar enhancements are found for the sample of very bright
(S $>$ 300 $\mu$Jy) radio sources (10/13, 77 \%), bright (S $\geq$ 70
$\mu$Jy) radio sources (28/45, 62 \%) and bright non-thermal sources
(16/31 = 52 \%). Essentially no enhancement is found for the faint (S
$<$ 70 $\mu$Jy; 8/64 = 12\%) sources. One possible interpretation is
that the faint sources, which are primarily non-thermal (i.e.,
$\alpha$ $<$ -0.4), arise from individual supernova remnants that are
too faint to be seen in the visible. This might be explained by the fact
that $\sim$10 \% of the UV light, and hence of the young star
formation responsible for producing most of the supernovae, is
associated with the field rather than the clusters.

3. A sample of 37 sources have been  studied in detail, showing correlations
between radio flux  and a variety of optical characteristics,
including cluster brightness, H$\alpha$ flux and equivalent width, 
A$_V$ extinction, and cluster ages.
The strongest correlation (7.9 $\sigma$) is between
the radio flux  and H$\alpha$ flux (Figure 12). This relationship
is probably the underlying cause behind several of the other correlations.
The most luminous objects, in both the radio and optical, are young
clusters with ages in the range 0 - 4 Myr and extinctions that range
from A$_V$ = 0.5 to 7.6.  In particular, the brightest radio source
is also the intrinsically brightest optical cluster (WS80 = 1139). In
addition, it is the strongest CO source as well as the strongest 15
micron source in the entire system. Furthermore, it has the highest
extinction and is one of the youngest clusters in our sample of 37
radio-optical matches. 

4. The radio bright phase lasts $<$ 10 Myr in our sample of clusters
with radio-optical matches, as demonstrated by the fact that all 36
objects have strong H$\alpha$ (i.e., log H$\alpha$ $>$ 1.5). This is
consistent with the interpretation that most of the radio emission
originates from hot thermal gas in compact HII regions. The gas is
ionized by O and B stars which have lifetimes $<$ 10 Myr, hence the
young ages for the radio sources.  It appears that, on average, a 6
Myr old cluster has been able to disperse enough dust around it to
reduce the value of A$_V$ to about 1 mag.

5.  The identification of WS80 with the strongest radio, CO, and IR
sources led to the expectation that many of the other very red sources
may have similar properties.  However, only two of the 37
radio-optical matches have $V-I$ $>$ 2.0.  In contrast, of the eleven
intrinsically brightest optical clusters in the overlap region with log
H$\alpha$ $>$ 3, we find eight of the clusters are strong radio
sources.  Hence, while there is nearly a one-to-one correspondence
between bright young clusters and radio sources, the very red clusters
are $\it not$ preferentially radio sources, contrary to earlier
suggestions.

6. As part of this project we developed a hybrid technique for
determining the ages of young star clusters, using a combination of $UBVI$ photometry
and H$\alpha$ equivalent width to break the age-reddening
degeneracy. We find that the Bruzual-Charlot (2000) models using the 
Padova spectral evolution tracks fit the
data reasonably well, while the Starburst99 (Leitherer \etal\/, 1999) models using the
Geneva tracks have an additional ``red loop'' that is inconsistent with
the data.

\bigskip
\bigskip

This work was supported by NASA grants GO-05416.01-93A and GO-07468.01-A.
We thank Eddie Bergeron for the suggestion to use the DSS image to
check the position of the WFPC2 image, and for help with the
probability calculations. We thank Bernhard Brandl for raising the
question of whether the radio sources have optical counterparts during
IAU symposium \# 207, since this led us to look into this question
more closely, resulting in this publication.
We also thank Adrienne Lancon for a  useful discussion, and 
Francois Schweizer, Michael Fall, and Bernhard Brandl for comments on the paper. Finally, we thank an anonymous referee for several useful comments.

\clearpage

\noindent {\bf APPENDIX A - Notes to Table 2}

\noindent 2-2: There are several clusters near or within the search
radius. Adopted 1161 ($V$ = 21.52, $V-I$ = 0.83, R = 0.48$''$) since
it was closer to the radio position than the brighter cluster 1143
($V$ = 21.14, $V-I$ = 0.76, R = 0.69$''$).

\noindent 2-4: There are several very bright clusters nearby. Adopted
1298 since it was the brightest ($V$ = 19.07, $V-I$ = 0.67, R = 0.32$''$).

\noindent 4-2: On the edge of region C (see Whitmore \etal\/
1999). Adopted the brightest object (2002, $V$ = 19.06, $V-I$ = 0.14), which is 0.66
$''$ from the radio position. The closest object was 2033 ($V$ = 20.96, $V-I$ =
0.33, R = 0.19$''$).

\noindent 4-5: On the edge of region D. Adopted the brightest object
(2410, $V$ = 18.70, $V-I$ = 0.13), which is 0.49$''$ from the radio
position. The closest object was 2399 ($V$ = 21.69, $V-I$ = -0.04, R = 0.30$''$).

\noindent 4A-16: There are three bright clusters in the
circle. Adopted the brightest object (5105 $V$ = 20.58, $V-I$ = 0.52),
which is 0.29 $''$ from the radio position.

\noindent 4A-9: There are several bright clusters just outside the
circle, including 3772 ($V$ =20.58, $V-I$ = 1.39; R = 0.81$''$).

\noindent 4B-2: Adopted the brightest cluster (3069, $V$ = 20.92, $V-I$ =
0.21), which is 0.46 $''$ from the radio position. The closest object
was 3003 ($V$ = 23.2, $V-I$ = 0.81, R = 0.45$''$).

\noindent 5-5: There are two good candidates in the circle. Adopted
the redder object which is slightly closer to the radio position
(7894, $V$ = 20.74, $V-I$ = 0.99, R=0.26$''$), but the radio source is probably a
combination with 7811 ($V$ = 19.38, $V-I$ = 0.30, R = 0.42$''$).

\noindent 7-2: Centered on part of the dust lane near the nucleus of NGC
4038. Adopted the very bright cluster 9089 ($V$ = 18.51, $V-I$ = 0.07) which
is just outside the search radius (R = 0.71$''$), but the
true source may be embedded.  The closest object was 9182 ($V$ = 22.10,
$V-I$ = 0.88, R = 0.31$''$).

\noindent 9-4: There are several candidates in the circle. Chose the
brightest cluster (5047, $V$ = 21.59, $V-I$ = 0.26, R = 0.25$''$).

\noindent 10-1: There are several candidates in the circle; the radio
source may be the sum of several of them. Adopted the brightest
cluster (5981, $V$ = 19.63, $V-I$ = 0.27, R = 0.36$''$).

\noindent 11-2: Near the very bright region S (8191, $V$ = 17.60, $V-I$ =
0.30, R=1.08$''$), but have adopted object 6665 ($V$ = 20.04, $V-I$ =
-0.15, R = 0.40$''$) since it is much closer to the radio position and
is a stronger H$\alpha$ source.

\noindent 13-7: Two roughly equal candidates. Adopted 11623 ($V$ = 20.82,
$V-I$ = 0.63, R = 0.46$''$) since it is slightly brighter than 11584
($V$ = 21.20, $V-I$ = 0.73, R = 0.48$''$).

\clearpage

\parskip 6pt plus 3pt minus 2pt

\noindent {\bf REFERENCES}

\noindent Biretta, J. et al.  2001, WFPC2 Instrument Handbook,
Version 6.0 (Baltimore: STScI), 210

\noindent Bruzual A. G., \& Charlot, S. 1996, private communication

\noindent Bruzual A. G., \& Charlot, S. 2000, private communication

\noindent Harris, J., Calzetti, D., Gallagher, J. S., Conselice, C. J., \& Smith, D. A. 2002, AJ, 122, \\ 
\phantom{aaaaa}3046

\noindent Kunze D. \etal\/ 1996, A\&A 315, L101

\noindent Leitherer, C., Schaerer, D., Goldader, J. D., Gonz\'alez Delgado, 
    R. M, Robert, C., Foo Kune, \\ 
\phantom{aaaaa}D., de Mello, D. F., Devost, D., \& Heckman,
      T. M. 1999, ApJS, 123, 3 (Starburst99)

\noindent Mathis J. S. 1990, ARA\&A, 28, 37

\noindent Mengel, S., Lehnert, M. D., Thatte, N., Tacconi-Garman, L. E. \& 
Genzel, R. ApJ, 2001, \\ 
\phantom{aaaaa}550, 280.

\noindent Mirabel, L, F., Vigroux, L., Charmandaris, V., Sauvage, M., Gallais, P., Tran, D., Cevarsky, \\ 
\phantom{aaaaa}C., Madden, S. C., Duc, P.-A. 1998 A\&A 333, L1

\noindent Neff, S. G. \& Ulvestad, J. S. 2000, AJ, 120, 670

\noindent Stiavelli, M., Panagia, N., Carollo, M., Romaniello, M. Heyer, I., Gonzaga, S. 1998, ApJL, \\ 
\phantom{aaaaa}492, L135

\noindent Toomre, A. 1977, in The Evolution of Galaxies and Stellar
Populations, edited by B.M. \\ 
\phantom{aaaaa}Tinsley and R. B. Larson (Yale University
Press, New Haven), p. 401

\noindent Vigroux, L. \etal\/ 1996, A\&A, 315, L93

\noindent Voit, M. 1997, HST Data Handbook (Baltimore: STScI)

\noindent Young, J. S., Allen, L.,
 Kenney, J. D. P., Lesser, A. \&
 Rownd, B.1996, AJ, 112, 1903

\noindent Whitmore, B. C., Miller, B. W., Schweizer, F., \& Fall, S. M. 1997, 
AJ, 114, 1797

\noindent Whitmore, B. C., \& Schweizer, F. 1995, AJ, 109, 960

\noindent   Whitmore, B. C., Zhang, Q., Leitherer, C., Fall, S. M.,  Schweizer, F. \&
\& Miller, B. W.\\ 
\phantom{aaaaa}1999, AJ, 118, 1551 

\noindent Wilson, C. D., Scoville, N., Madden S. \& Charmandaris, V. 2000, ApJ, 542, 120

\noindent Zhang, Q., Fall, M. \& Whitmore, B. C.  2001, ApJ, 561, 727

\clearpage 

\centerline{\bf FIGURE CAPTIONS}

\noindent Figure 1 - Positions of the strong (S $\geq$ 70 $\mu$Jy)
thermal ($\alpha$ $\geq$ --0.4) radio sources (Neff \& Ulvestad 2000)
superposed on the F814W HST image of the overlap region.  The top
panel  shows the original positions while the top panel (1b) shows
the positions when the HST image is moved 1$\farcs$2 toward the
southwest. An area in the upper right (Knot B) is shown with a factor
of 10 dimunization in Figure 1b in order to show more detail in a
region that would otherwise be saturated.  

\noindent Figure 2 - The probability of having a given number of
matches for 13 trials. The histogram shows the results of a
Monte-Carlo simulation, as described in the text. The curve is a
binomial probability determined by the mean number of predicted
matches. The number of matches using the original positions (3) is
consistent with being from a random distribution while the 11 matches
resulting after the 1$\farcs$2 offset has been made has a probability
of 4.6 $\times$ 10$^{-7}$, based on the binomial probability.

\noindent Figure 3 - Locations of all radio sources from Neff \&
Ulvestad (2000) superposed on the HST F814W image (shifted 1$\farcs$2
to the southwest).  The circles are thermal sources ($\alpha$ $>$
--0.4); the squares are non-thermal sources ($\alpha$ $<$
--0.4). Large symbols are for the bright sources (S $>$ 70); small
symbols are for the faint sources (S $<$ 70).

\noindent Figure 4 - Locations of radio sources in part of the western
loop, after the 1$\farcs$2 offset has been applied to the HST
images. The left panel (4a) is the F814W image, the right panel (4b)
is the H$\alpha$ image (with the F814W continuum subtracted).  The
large circles show the 0$\farcs$5 search radius around the radio
source. The small circles show the candidates from the
``brightness-selected'' sample while the small squares show the
candidate from the ``H$\alpha$-selected sample. Note that the very
bright cluster near the top of Figure 4a (Knot S; near cluster candidates 8064 
and 8086) is not a radio
source, but the region of strong H$\alpha$ below Knot S (Figure 4b) is
a radio source.

\noindent Figure 5 - $U-B$ vs $V-I$ diagram with data points from the 37
optical counterparts (middle panel) from the brightness-selected sample. The solid curve
is the Bruzual-Charlot (2000) model using theoretical isochrones; the
dot-dashed curve is the Bruzual-Charlot (2000) model using empirical
isochrones; and the dotted curve is the Starburst99 (Leitherer \etal\/, 1999)
model, all for solar metallicity.  The straight dashed lines show the
reddening vectors for ages 1, 5, and 10 Myr from the Bruzual-Charlot
(2000) models with theoretical isochrones and Mathis (1990) reddening
law. Various clusters are
identified and discussed in the text. The bottom panel
shows the corresponding diagram for the 100 brightest (apparent) clusters
in the Antennae. Note that there are almost no clusters in the
region of the diagram that would be populated by the
``red loop'' in the Starburst99 models (i.e., ages 8 -- 13 Myr).

\noindent Figure 6 - log H$\alpha$ vs. log Age. The curve
shows a Starburst99 (Leitherer \etal\/, 1999) model for an instantaneous burst,
solar metallicity model (i.e.,
their Figure 45). The adopted ages are the ``best guess'' values, as discussed
in the text.

\noindent Figure 7 - $B-V$ vs $V-I$ diagram with data from the 37 optical
counterparts in the brightness-selected sample. The curve shows the Bruzual-Charlot (2000) model using
theoretical isochrones; the dashed lines show the Mathis (1990)
reddening vectors for ages 1, 5, and 10 Myr, which are nearly parallel
with the age models.  The ten clusters with the largest values of
H$\alpha$ are shown using open circles, demonstrating how the
presence of emission lines (4861, 4959, 5007 \AA) in the $V$ passband
affect the broadband color determinations.

\noindent Figure 8 - $B-V$ residuals from the 1 Myr reddening vector in
Figure 7 as a function of log H$\alpha$. The dashed line shows the
curve (Equation 2) used to correct for the presence of emission lines in the V
passband for the subsequent analysis. See text for details.

\noindent Figure 9 - The four figures used to estimate the ``best
guess'' age estimates for the clusters. $B-V$(cor) and $V-I$(cor) have been corrected for
the presence of emission lines in the $V$ passband, as
discussed in the text. The dashed lines show the Mathis (1990)
reddening vectors for ages 1, 4.8 and 8.7 Myr. 
The open circle show the data for cluster 5105. While multiple
ages are possible based on the color-color diagrams, the
strong H$\alpha$ strength indicates that an age $\sim$3 Myr
is the correct answer.

\noindent Figure 10 - Eight plots showing the correlations between
various radio and optical properties for the brigthness-selected
sample. The solid circles are for sources with $\alpha > -0.8$ while
the open circles are for sources with $\alpha < -0.8$ (i.e., 
non-thermal). The significance of each correlation is shown in the
upper right corner of the panel. Obvious outlyers (labeled) have been removed
from Figures 10d and 10g before making the fits.

\noindent Figure 11 - The log S vs. log H$\alpha$(EW) diagram for the three
samples described in the text. Note that the scatter is much
smaller in the radio-position sample, suggesting that the primary
correlation is with H$\alpha$ flux rather than equivalent width.

\noindent Figure 12 - The log S vs. log H$\alpha$ (flux) diagram for
the radio-position sample. The resulting correlation (7.9 $\sigma$) is
stronger than between any other parameters for any of the three
samples. Hence, the relationship between the total H$\alpha$ (flux)
and radio flux (S) is probably the underlying correlation which is
responsible for many of the other trends (e.g., Figure 10a, 10b, 10e).

\noindent Figure 13 - $U, B, V, I, V/I,$ and H$\alpha$ images of the
regions around WS80 (cluster 1139) and WS355 (cluster 7086), two very red clusters identified by
Whitmore \& Schweizer (1995).  The contrast for the V/I image is
adjusted to show the blue objects as black and the red objects as
white.  Note that WS80 is a strong H$\alpha$ source while WS355 has
essentially no H$\alpha$ associated with it. This explains why WS355
is not a radio source.

\begin{figure*}
\includegraphics{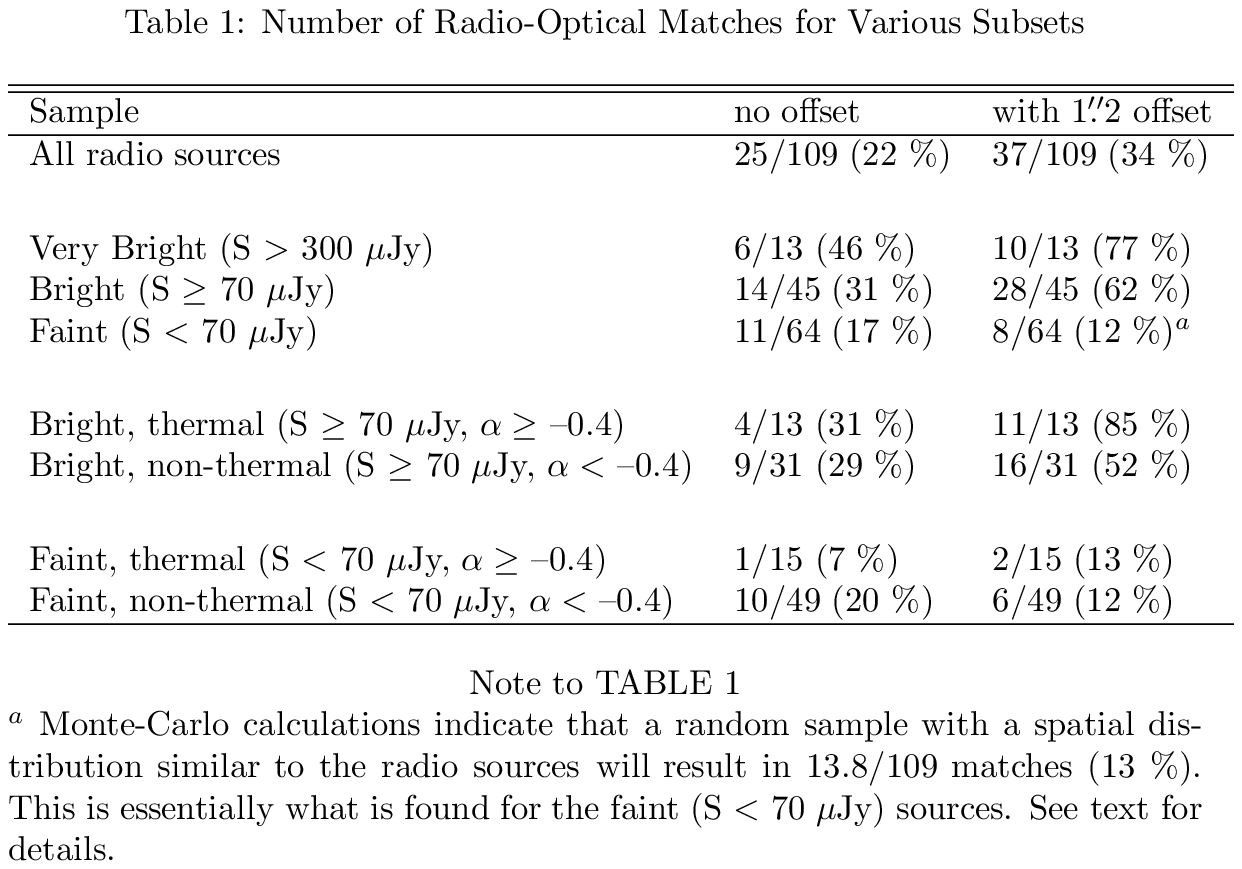}
\vspace{19.0cm}
\end{figure*}

\begin{figure*}
\includegraphics{whitmore2_table2.ps}
\vspace{19.0cm}
\end{figure*}

\begin{figure*}
\includegraphics{whitmore2_table3.ps}
\vspace{19.0cm}
\end{figure*}

\begin{figure*}
\includegraphics{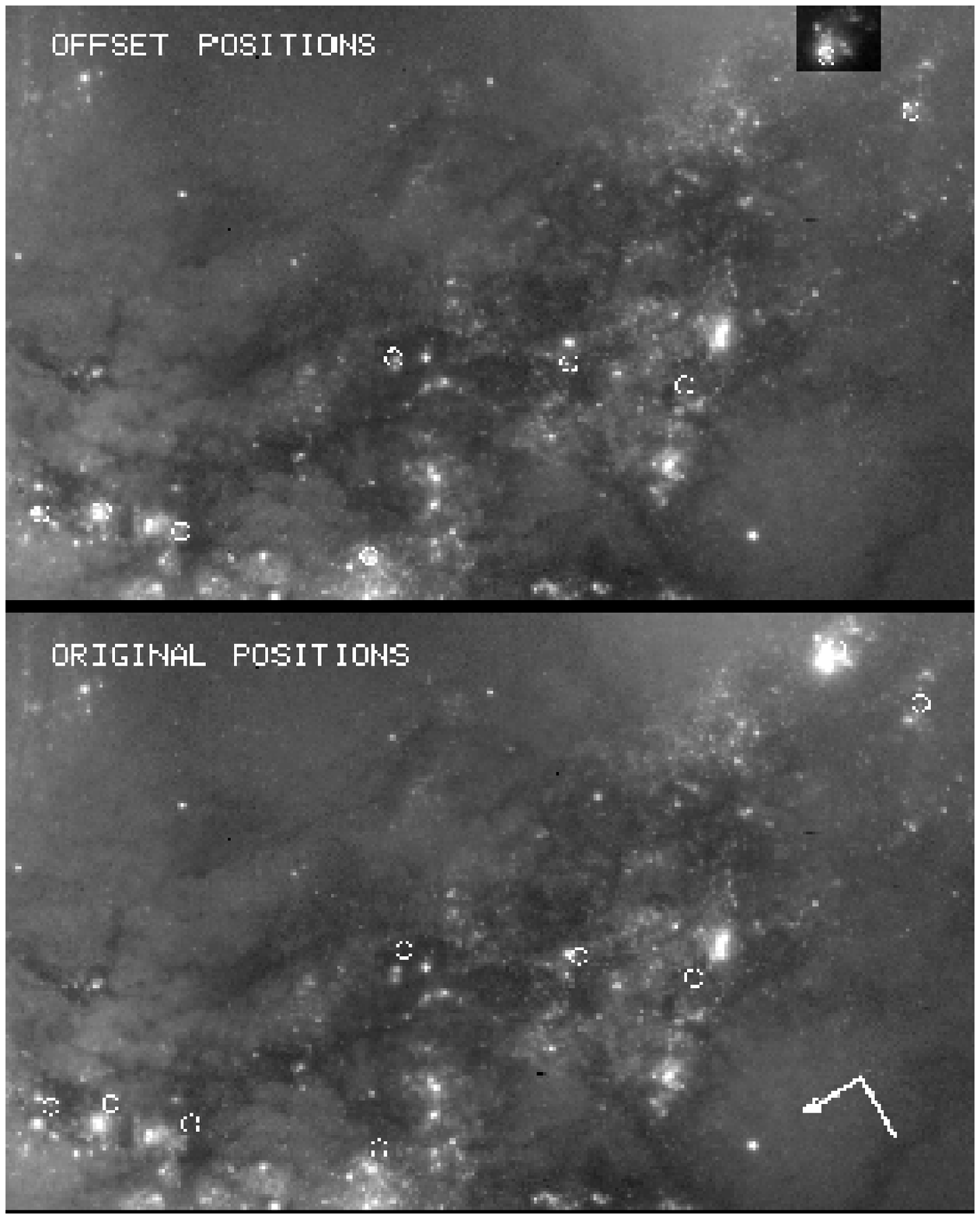}
\vspace{19.0cm}
\caption{}
\end{figure*}

\begin{figure*}
\includegraphics{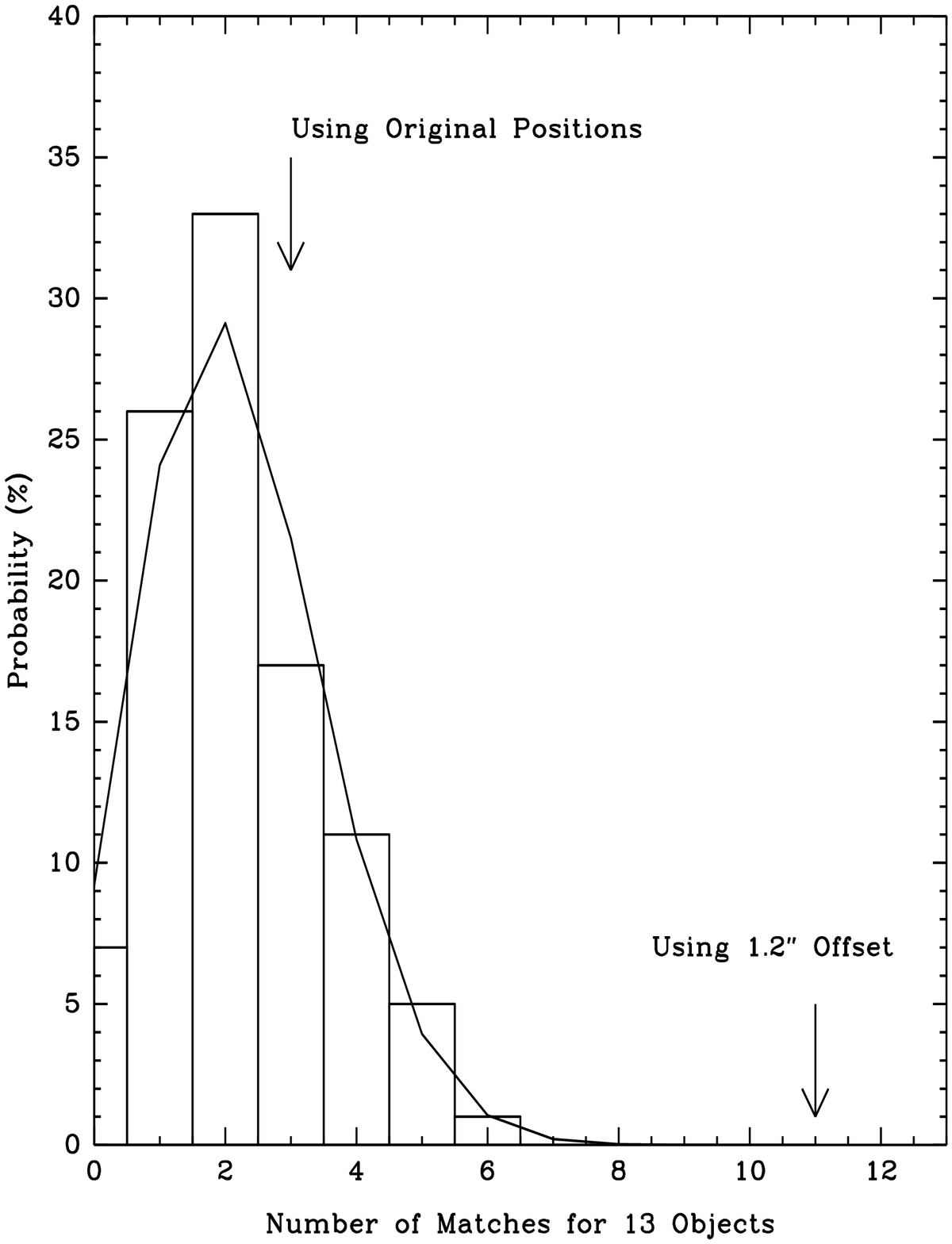}
\vspace{17.0cm}
\caption{}
\end{figure*}

\begin{figure*}
\includegraphics{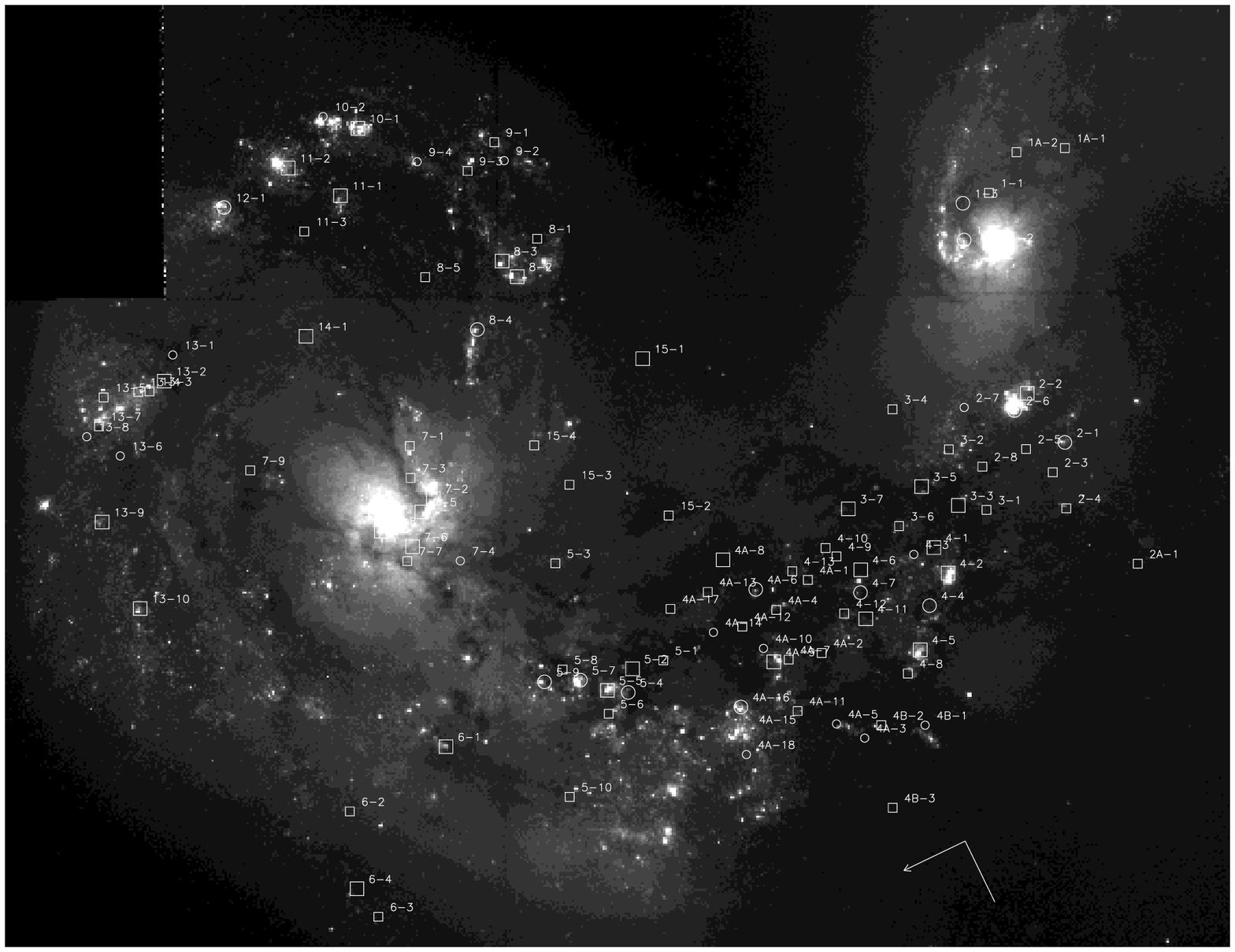}
\vspace{19.0cm}
\caption{}
\end{figure*}

\begin{figure*}
\includegraphics{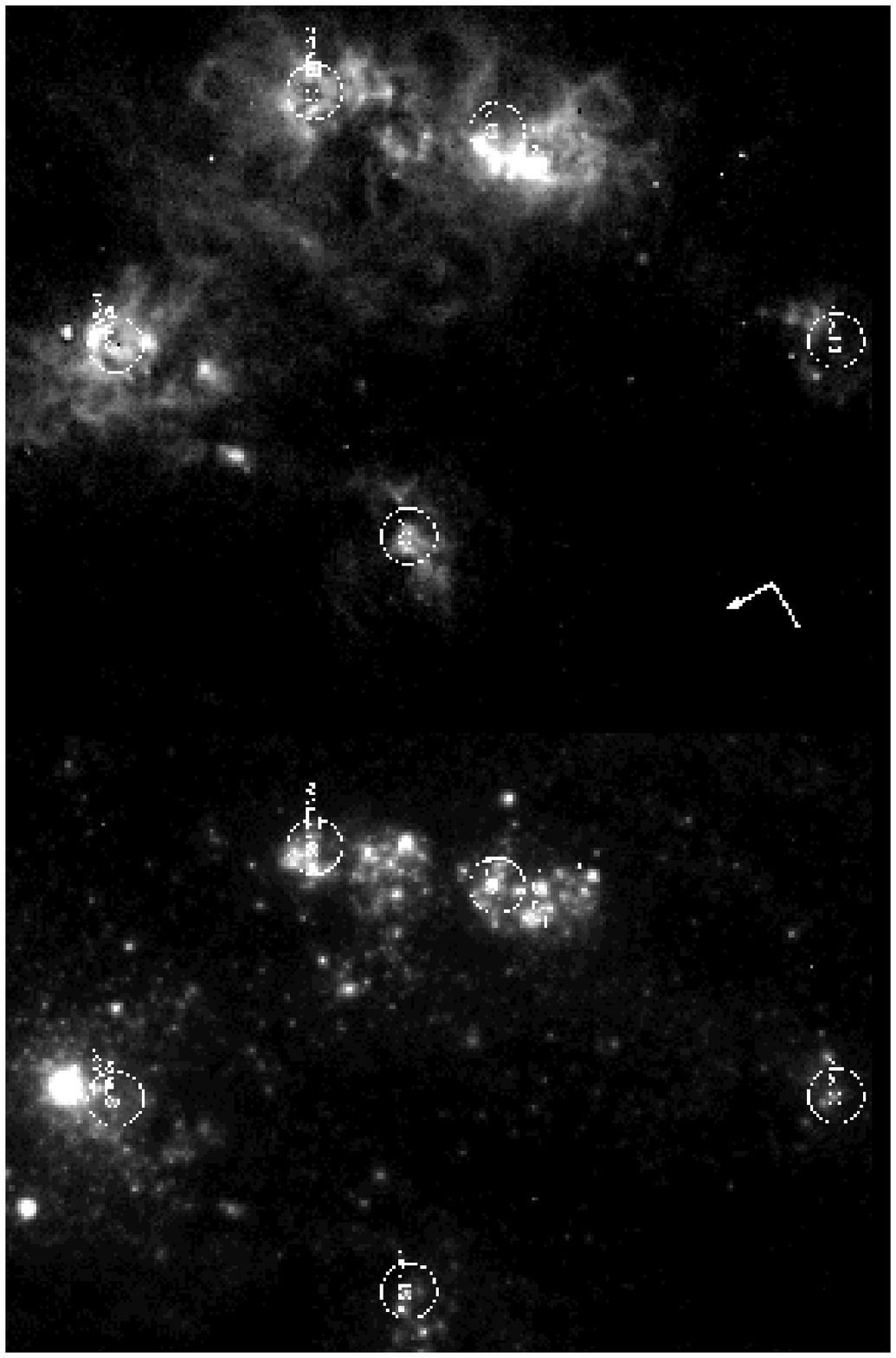}
\vspace{17.0cm}
\caption{}
\end{figure*}

\begin{figure*}
\includegraphics{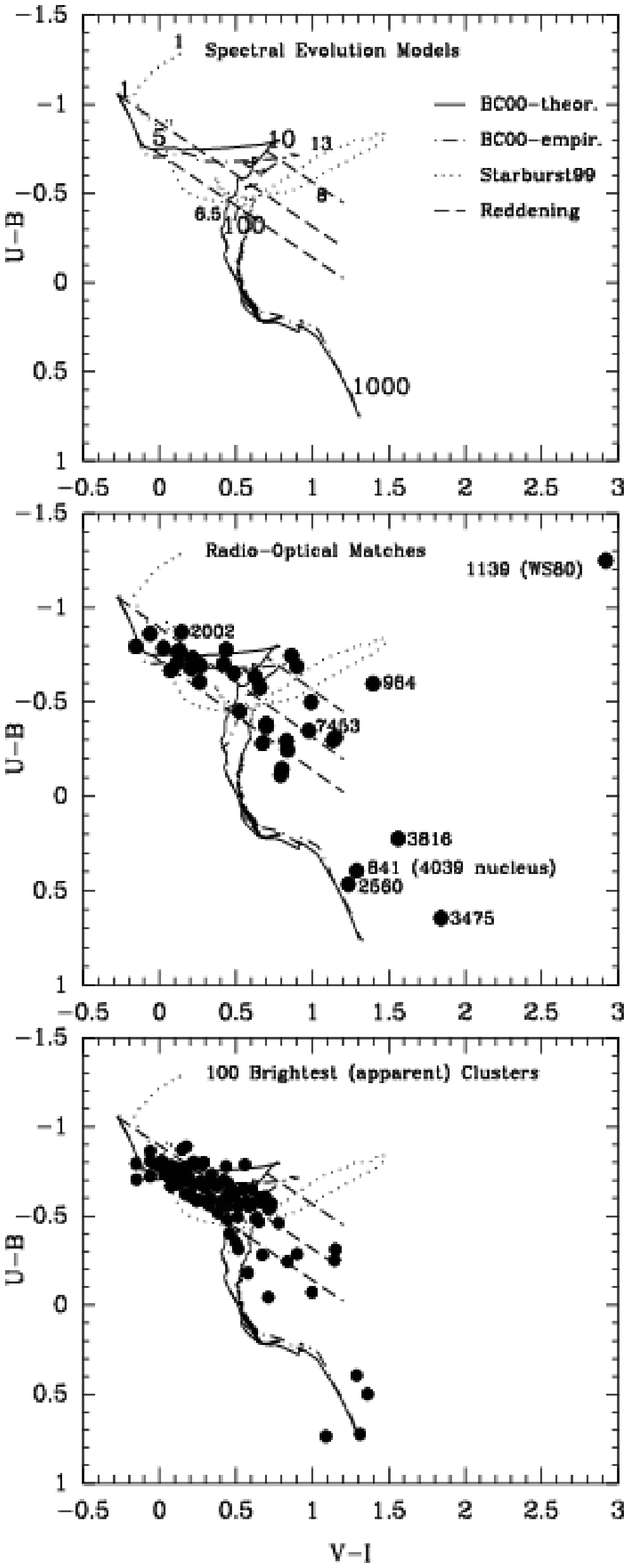}
\vspace{18.0cm}
\caption{}
\end{figure*}

\begin{figure*}
\includegraphics{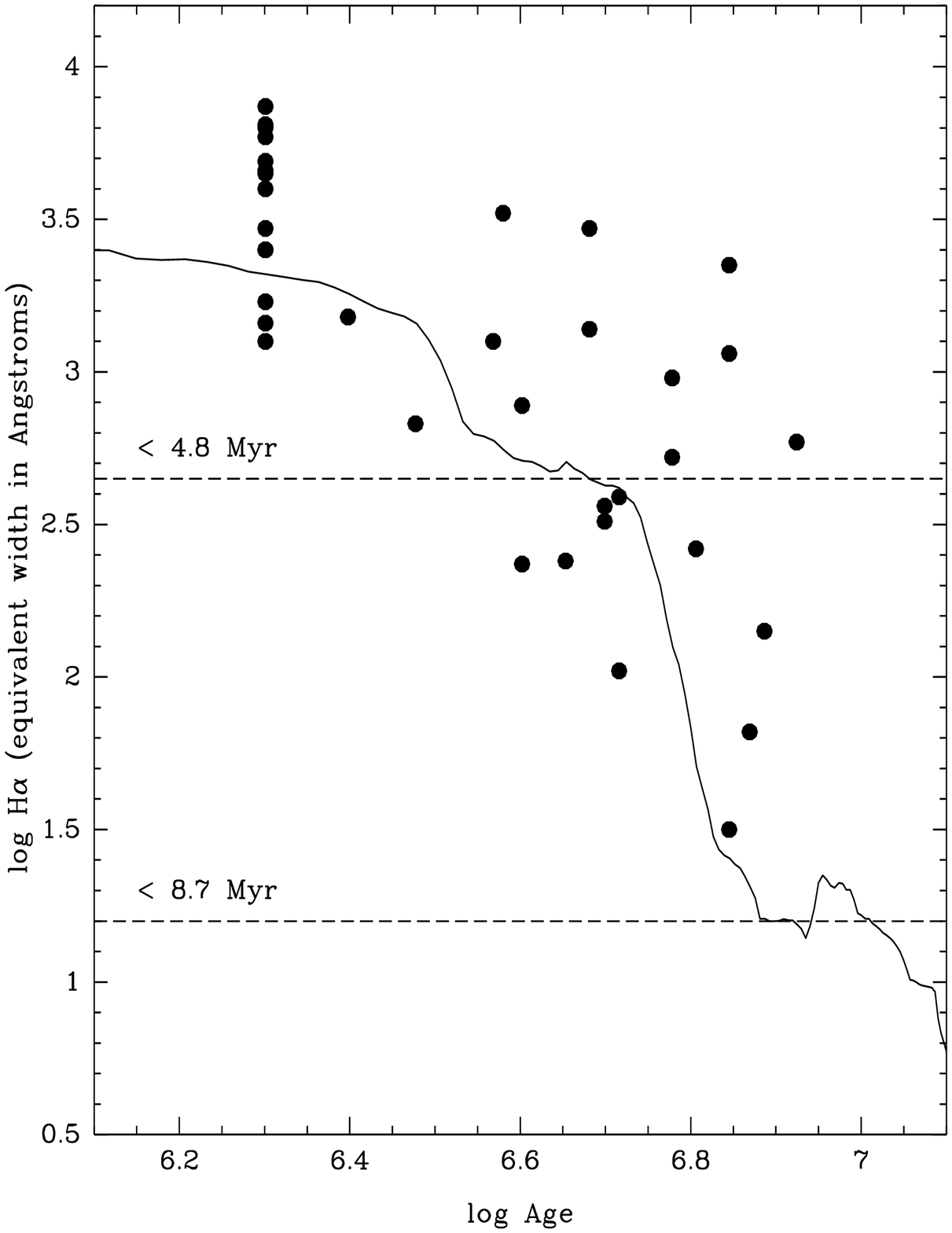}
\vspace{17.0cm}
\caption{}
\end{figure*}

\begin{figure*}
\includegraphics{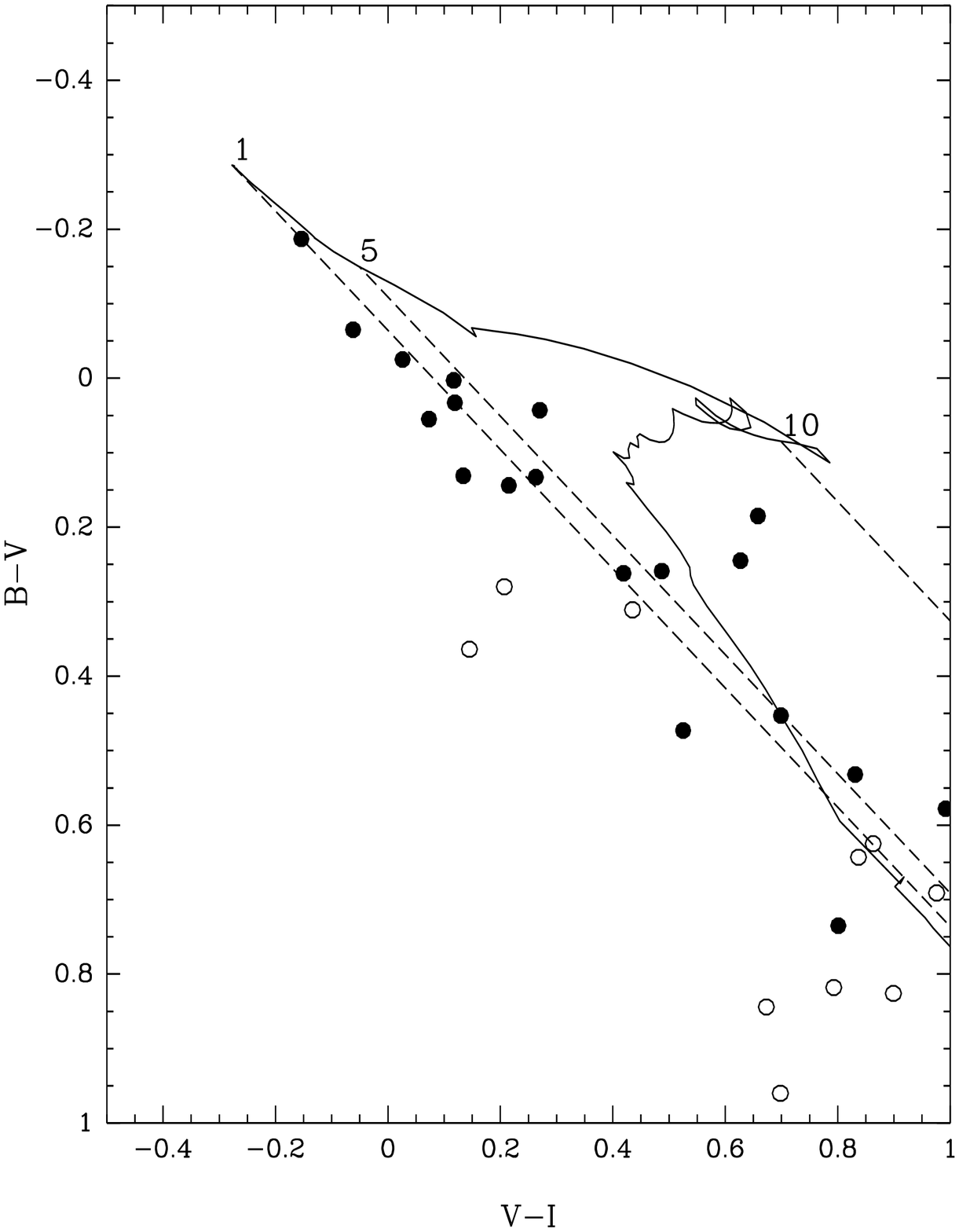}
\vspace{17.0cm}
\caption{}
\end{figure*}

\begin{figure*}
\includegraphics{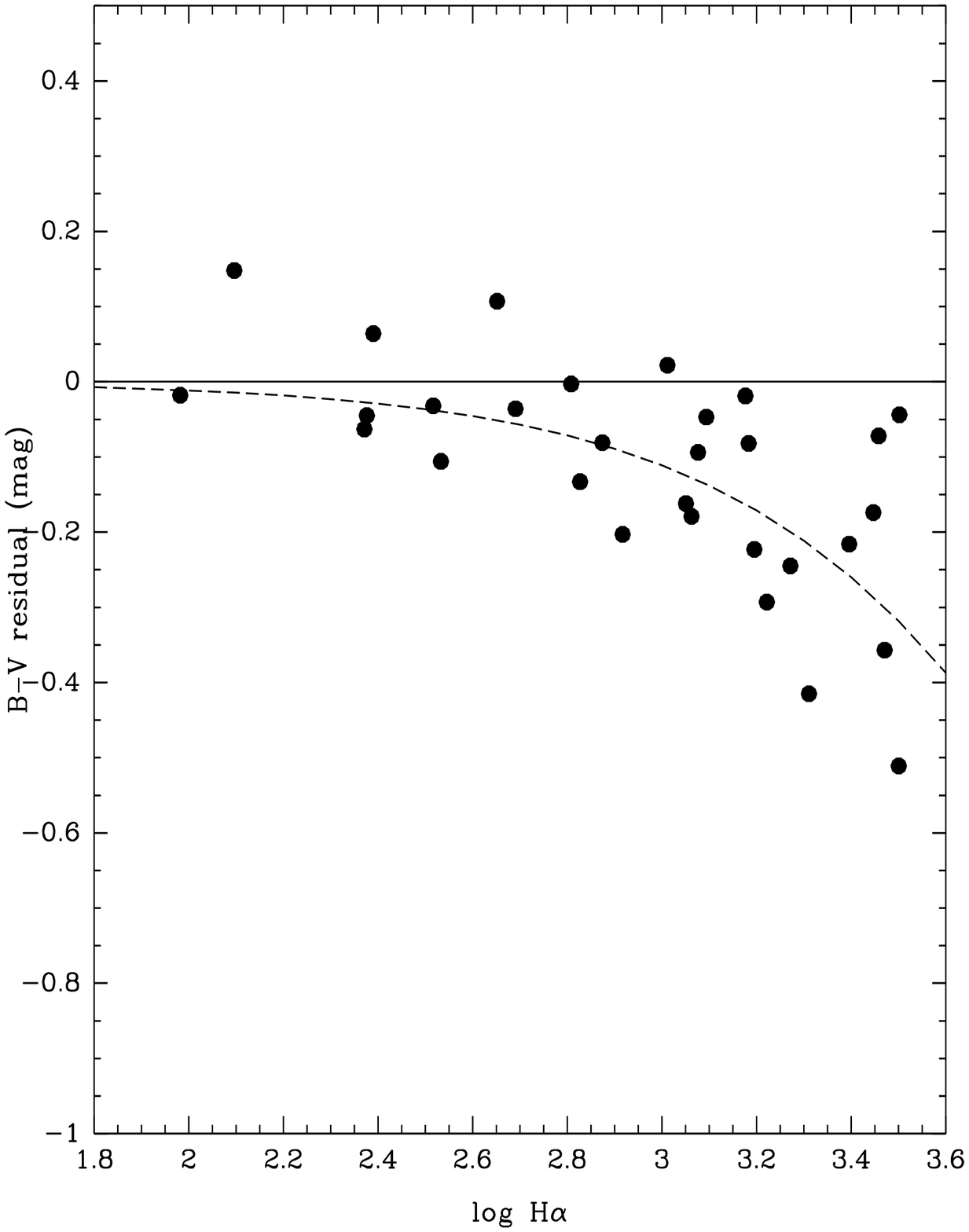}
\vspace{17.0cm}
\caption{}
\end{figure*}

\begin{figure*}
\includegraphics{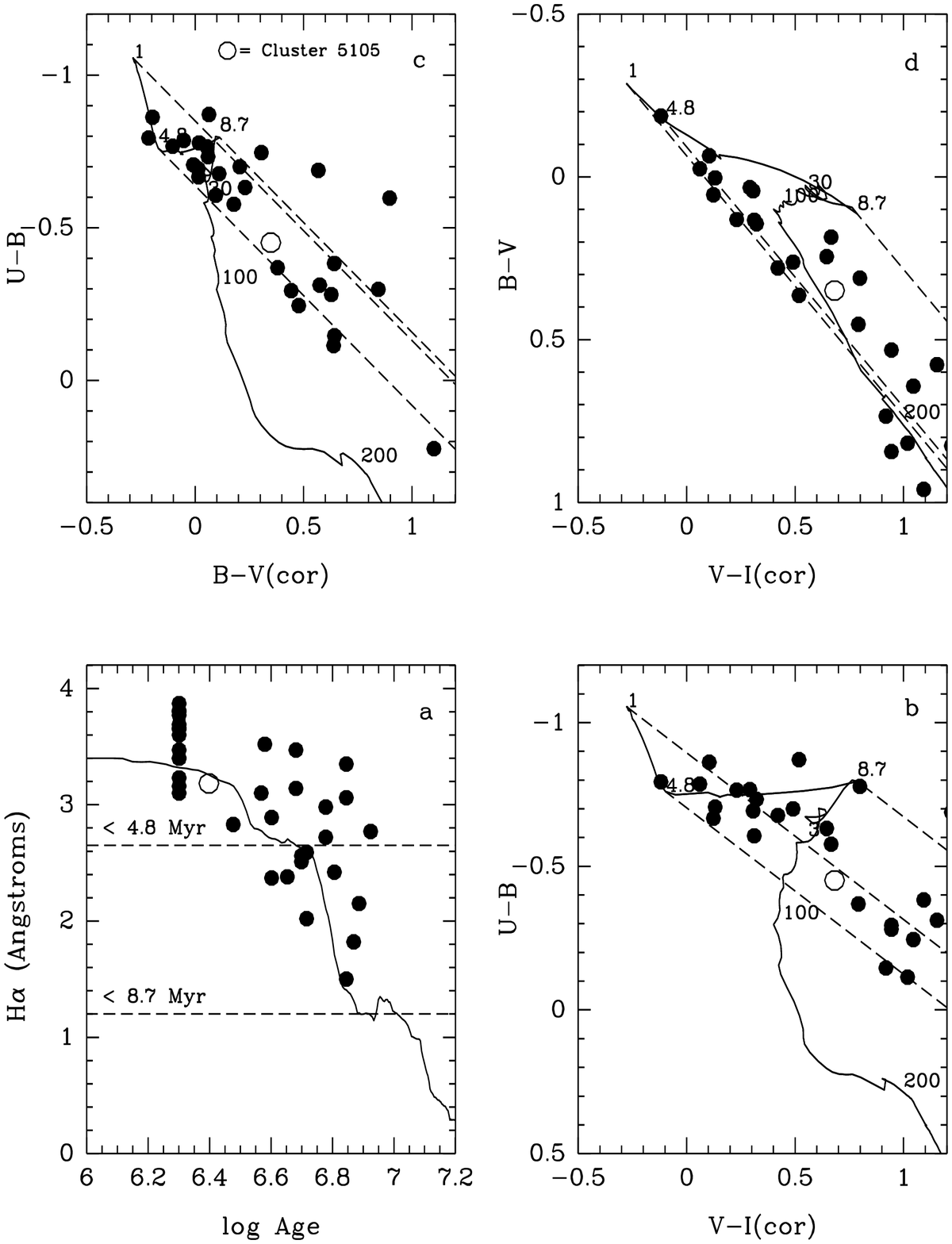}
\vspace{17.0cm}
\caption{}
\end{figure*}

\begin{figure*}
\includegraphics{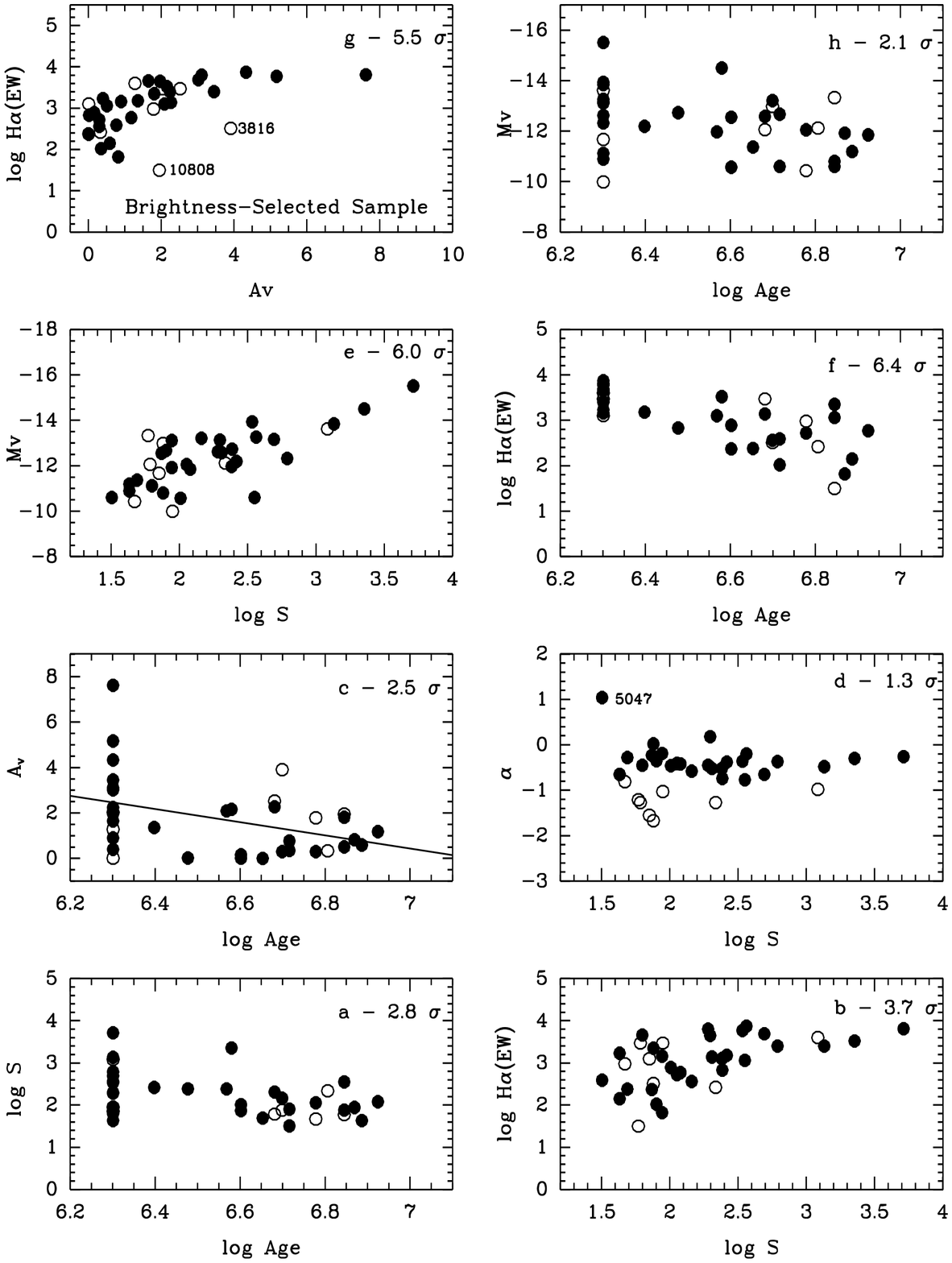}
\vspace{17.0cm}
\caption{}
\end{figure*}

\begin{figure*}
\includegraphics{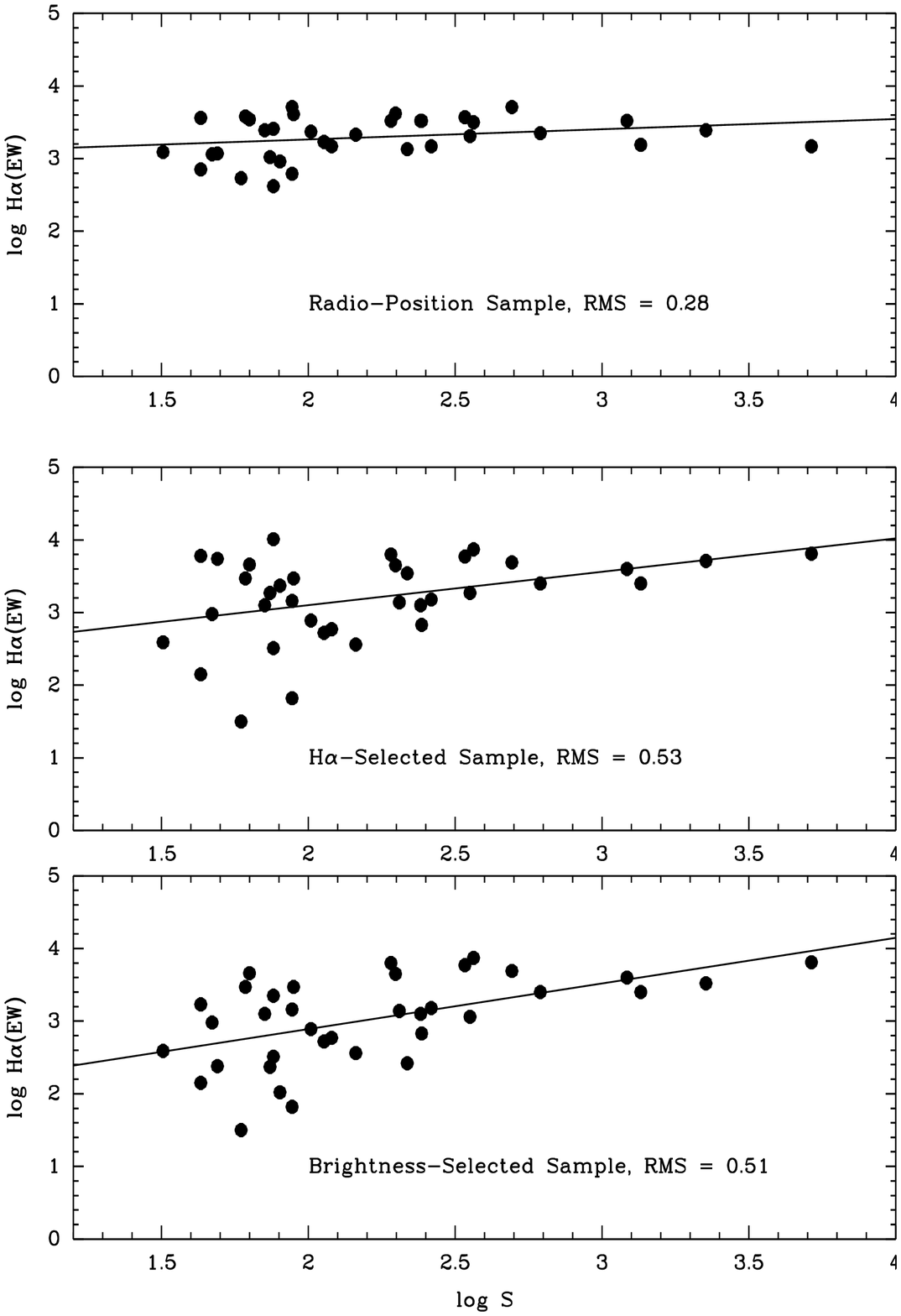}
\vspace{17.0cm}
\caption{}
\end{figure*}

\begin{figure*}
\includegraphics{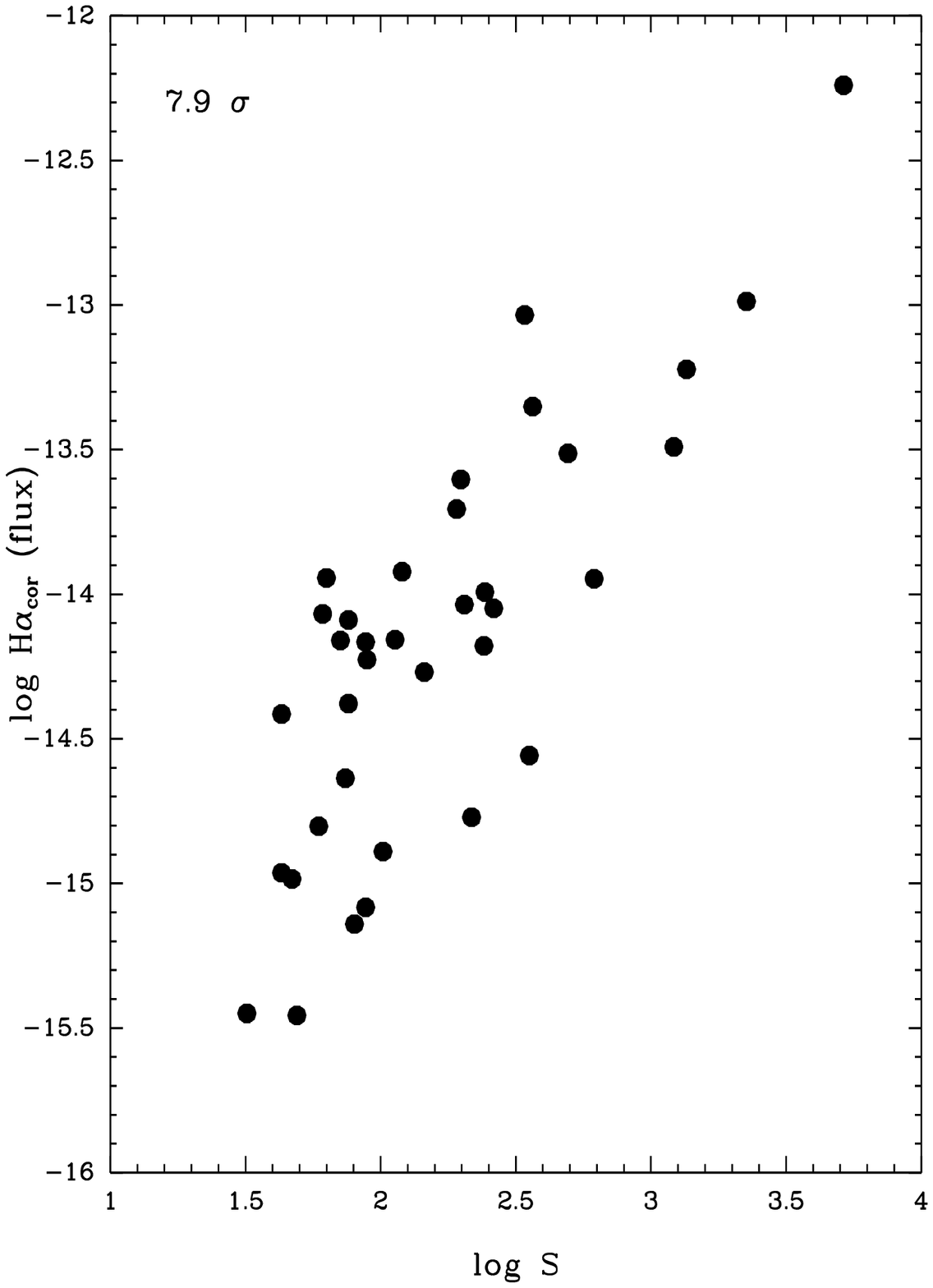}
\vspace{17.0cm}
\caption{}
\end{figure*}

\begin{figure*}
\includegraphics{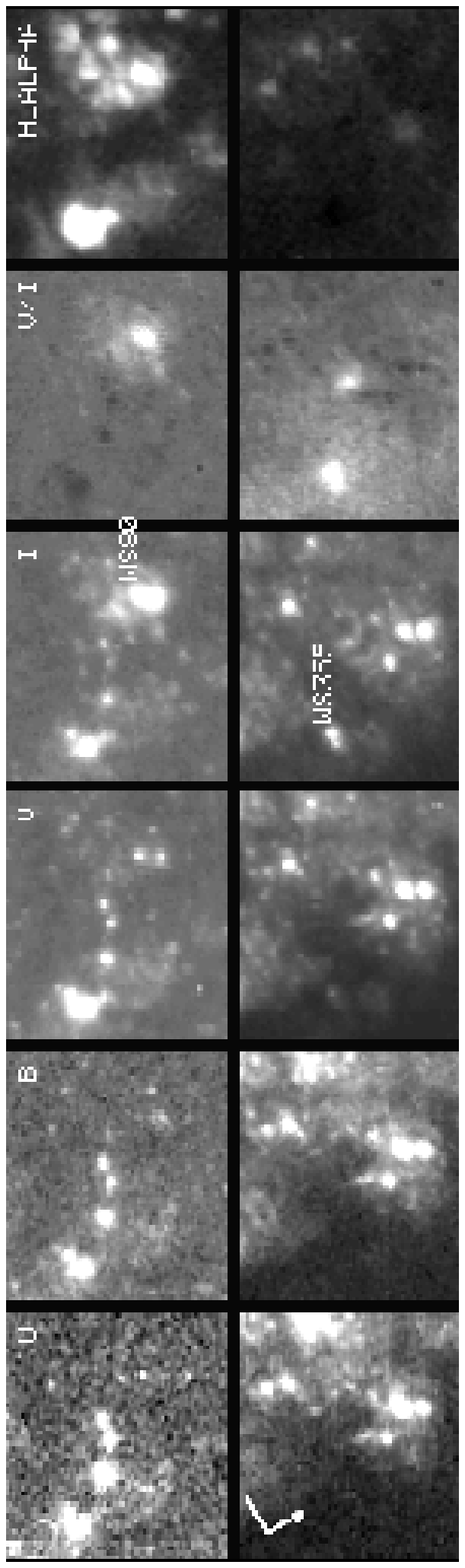}
\vspace{17.0cm}
\caption{}
\end{figure*}

\end{large}			

\end{document}